\documentclass[11pt,preprintnumbers,aps,amssymb,nofootinbib,amsmath]
{revtex4}
\usepackage{epsfig,epsf}
\usepackage{bm} % puts greek and math symbols in boldface using \bm
\newcommand{\beq}{\begin{equation}}
\newcommand{\beql}[1]{\begin{equation}\label{#1}}
\newcommand{\eeq}{\end{equation}}
\newcommand{\bea}{\begin{eqnarray}}
\newcommand{\eea}{\end{eqnarray}}
\newcommand{\eq}[1]{(\ref{#1})}
\newcommand{\fig}[1]{Fig.~\ref{#1}}
\renewcommand{\sec}[1]{Sec.~\ref{#1}}
\newcounter{topiccounter}
\renewcommand{\b}[1]{\mathbf{#1}}
\newcommand{\im}{\mathrm{Im}\,}
\newcommand{\real}{\mathrm{Re}\,}

\begin{document}

\preprint{RBRC-786}

\title{Multi-photon interactions in lepton photo-production on nuclei at high energies}

\author{Kirill Tuchin$\,^{a,b}$\\}

\affiliation{
$^a\,$Department of Physics and Astronomy, Iowa State University, Ames, IA 50011\\
$^b\,$RIKEN BNL Research Center, Upton, NY 11973-5000\\}

\date{\today}


\begin{abstract}

We study multi-photon effects in quantum electrodynamics  in lepton photo-production on heavy nuclei and nuclear medium at high energies. We are interested in energy, charge density and nuclear geometry dependence of the cross sections. We use the impact parameter  representation  that allows  us to reduce the problem of photo-production to the problem of propagation of electric dipoles in the nuclear Coulomb field. In the framework of the Glauber model  we resum an infinite series of multiphoton amplitudes corresponding to multiple rescattering of the electric dipole on the nucleus.  We find that unitarity effects arising due to multi-photon interactions are small and energy-independent  for scattering on a single nucleus, whereas in the case of macroscopic nuclear medium they saturate the geometric limit of the total cross section.  We discuss an analogy between nuclear medium and intense laser beams.

\end{abstract}

\maketitle

%%%%%%%%%%%%%%%%%%%%%%%%%%%%%%%%%%%%%%%%
\section{Introduction}\label{sec:intr}

In this article we offer a new perspective on the multi-photon interactions in lepton photo-production on a heavy nucleus at high energies.  Multi-photon effects arise due to multiple rescattering of the projectile photon on the protons of the nucleus. Each scattering contributes a factor of $\alpha Z$ to the cross section and must be taken into account for heavy nuclei. Traditionally, this is accomplished by solving the Dirac equation in the external Coulomb field of the nucleus. An explicit assumption of this approach is that the nucleus can be treated as a point-like particle. The goal of this article is to solve the photo-production problem in a general case, by explicitly taking the nuclear size into account. This allows us to evaluate  the finite nuclear size effects in lepton photo-production. However, our main interest is to use the developed formalism to investigate the photo-production in another extreme case -- inside the nuclear medium (e.g.\ a thin film).  We will argue that the nonlinear effects are very different in these two opposite cases. The formalism that we develop and use in this work can be also applied to the very interesting problem of  photo-production on intense laser beams \cite{Tuchin:2009ir}. There the beam size can be tuned so that both limits can perhaps be realized experimentally. 

Our approach to the multi-photon effects is based on the Glauber model \cite{Glauber:1987bb}, which allows a proper treatment of the nuclear geometry. In \sec{sec:Born} and the Appendix we show that the Glauber model calculation of the pair production cross section agrees with the traditional approach pioneered by Bethe and Maximon \cite{Bethe:1954zz}, which is based on a solution of the Dirac equation in an external Coulomb field (see \cite{Ivanov:1998dv,Ivanov:1998ru,Baltz:2001dp,Lee:2001ea,Bartos:2001jz,Baur:2007zz,Ivanov:1998ka} for recent developments). This is an important result of this paper. It opens the possibility of a unified approach to the geometric effects in QED. In particular, in this paper we study the dependence of the lepton pair photo-production on the density of charge sources  and nuclear geometry.

The paper is structured as follows. In \sec{dipole} we write the 
cross section in a mixed coordinate--momentum representation by transforming the scattering amplitude to the transverse coordinate space \cite{Cheng:1969eh,dip}. The cross section then becomes a convolution of the photon wave function, describing splitting of the photon into a lepton-antilepton pair characterized by the  transverse vector $\b r$, and the elastic amplitude $i\Gamma^{l\bar lZ}$ for the scattering of the lepton electric dipole $l\bar l$ off the nucleus $Z$ at some impact parameter $\b B$, see \eq{gammap}. The advantage of this representation becomes clear if we recall that the trajectory of an ultra-relativistic particle in an external field is a straight line. This implies that the dipole representation diagonalizes the scattering matrix. It was shown in \cite{Ivanov:1998ka} that the lepton photo-production amplitudes take a  rather simple form in the dipole representation.  This representation was used in \cite{Baltz:2006mz} for numerical study of the impact parameter dependence of the pair production. 

In \sec{glauber} we use the Glauber theory for multiple scattering \cite{Glauber:1987bb} to express the dipole--nucleus scattering amplitude $i\Gamma^{l\bar lZ}$ through the dipole--proton one $i\Gamma^{l\bar lp}$. This can be done if the nuclear protons can be considered as independent scattering centers. For large $Z$ their distribution thus follows the Poisson law that leads to a particularly simple expression for the amplitude \eq{aZ}. 
To avoid  possible confusion we emphasize, that we consider ultra-relativistic scattering on a ``bare" nucleus, i.e.\ a nucleus stripped of all electrons. Such nuclei are provided by the high energy heavy ion colliders such as RHIC and LHC. Therefore, the problem of screening of the nuclear Coulomb field by electrons -- which is an important issue at high energy photon-atom interactions -- is not relevant here.

We calculate the scattering amplitude for dipole--proton scattering in \sec{scat.ampl.}.  The momentum exchanged in the high energy collision is very small. The corresponding impact parameter is large. The maximal impact parameter $b'_\mathrm{max}$ increases with energy. For a single nucleus, it always holds that $b'_\mathrm{max}\gg R$. The region of large impact parameters is where the dipole cross section picks up its logarithmically enhanced energy dependence. Therefore, $b'_\mathrm{max}$ is the effective radius of electromagnetic (EM) interaction of the photon with the nucleus. This is very different from the strong interactions where the Yukawa potential drops exponentially at distances $\sim 1/(2m_\pi)$ much shorter than the nucleus radius. 

To make connection with the world of high photon densities, we  
 consider photo-production off a nuclear medium with transverse extent 
 $d\gg b'_\mathrm{max}$ and the longitudinal extent $L\ll d$. This can be thought of as a prototype of an intense laser pulse. Any photon can be converted into an electric dipole at the expense of an additional $\alpha$ which, however, impacts only the total rate, but not the dynamics of the multiple scatterings. This motivated us to consider the photo-production off the nuclear medium in more detail. Also, as has been mentioned, this case is similar in many aspects to the well-studied dynamics of the strong interactions.

At first,  in \sec{sec:Born} we study the Born approximation with the leading logarithmic accuracy. Born approximation corresponds to scattering on only one proton in a nucleus.  For  realistic nuclei, the photo-production cross section reduces to the Bethe-Heitler formula \eq{gZLO-2}\cite{Bethe:1934za} with a characteristic  $Z^2$ dependence on the proton number. In this case the elastic scattering amplitude $i\Gamma^{l\bar lp}$ is approximately real, see \eq{sa-electron}, corresponding to the dominance of elastic processes over the inelastic ones. This is not difficult to understand: a significant part of the cross section originates at distances much larger than the nuclear radius; the nuclear density there is small.  
On the contrary, in the case of the nuclear medium the amplitude $i\Gamma^{l\bar lp}$ is approximately imaginary, see \eq{z2},\eq{resmall3}.  This implies the dominance of inelastic processes. Formally, this happens because of an approximate cylindrical symmetry of the nuclear distribution around a small dipole on one hand, and anti-symmetry of the real part of the scattering amplitude $\real [i\Gamma^{l\bar lp}]$ with respect to inversion $\b b\to -\b b$, see \eq{llog}, on the other hand. For the nuclear medium, the photo-production cross section is proportional to $Z$ in the Born approximation.

Multiple rescatterings induce a correction to the Born approximation that generally depend on energy $\sqrt{s}$ and nuclear charge $Z$. These effects are studied  in \sec{resum}.
Expectedly, the multi-photon effects  lead to rather different expressions for the  resumed cross sections \eq{gammaZ-ee} and \eq{gammaZ-mm} for a single nucleus and for the nuclear medium.
For a single nucleus  the multi-photon processes resumed within the Glauber model do not depend on energy. In the Appendix, we demonstrate, using the method of  \cite{Ivanov:1998ka} that the Glauber approach reproduces the result of Bethe and Maximon, i.e.\ unitarity corrections beyond the leading logarithmic approximation \cite{Bethe:1954zz}.
In the opposite case of  the nuclear medium, the cross section grows as a function of energy and $Z$  until it saturates the black disk geometric limit. A problem of propagation of  positronium through matter was 
previously discussed by Lyuboshits and Podgoretsky in Ref.~\cite{Lyuboshits:1981ky} who reached similar conclusions.

A related problem has been recently addressed  by C.~Muller \cite{Muller:2008ys}   who considered scattering of an intense laser pulse on a heavy nucleus. He argues -- basing on work done with his collaborators \cite{Muller:2003zzd} -- that the resumation in the density of the laser beam can be phenomenologically significant if one studies the laser--nucleus interaction at the LHC.

%%%%%%%%%%%%%%%%%%%%%%%%%%%%%%%%%%

\section{Photo-production in the dipole model}\label{dipole}

%%%%
\begin{figure}[ht]
      \includegraphics{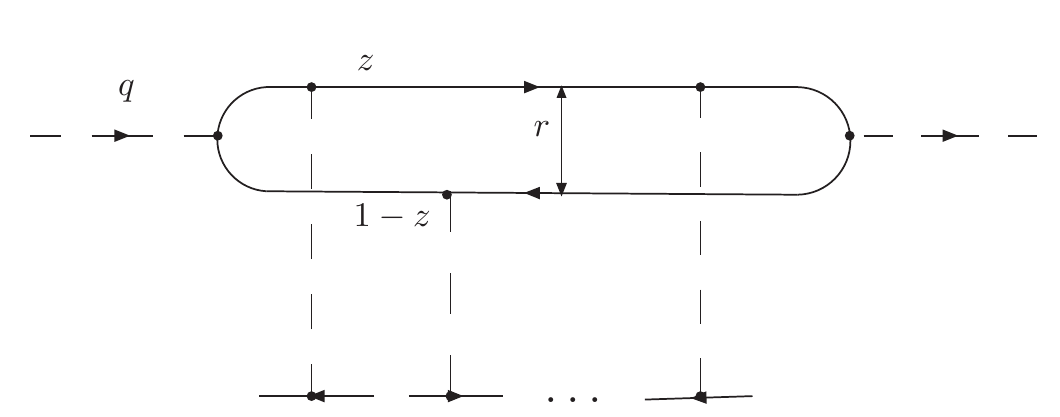} 
  \caption{Photo-production of a lepton pair. Dashed lines denote photons, solid lines denote fermions. The ellipses indicates summation over all proton numbers. Note that the $t$-channel photons can hook up to either fermion or anti-fermion lines. }
\label{fig:gamma}
\end{figure}
\begin{figure}[ht]
      \includegraphics[width=3cm]{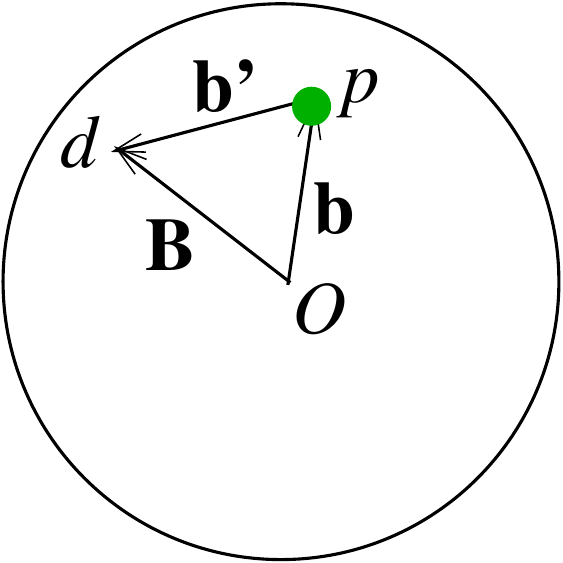} 
  \caption{Geometry of the dipole-nucleus collision in the transverse plane. $p$, $d$, and $O$ denote proton, center of mass of the dipole and the nucleus center positions correspondingly. $\b b'=\b B-\b b$, $b\le R$. }
\label{fig:nucl}
\end{figure}

%%%%%
One of the Feynman diagrams for the photo-production process is shown in \fig{fig:gamma}. In the high energy limit $s\gg -t$, where $s$ and $t$ are the usual Mandelstam variables,  the cross section for the diagram in \fig{fig:gamma} can be written as \cite{Nikolaev:1990ja}
\beq\label{gammap}
\sigma^{\gamma Z}_\mathrm{tot}(s)=\frac{1}{2}\int \frac{d^2r}{2\pi}\, \int_0^1 d\zeta\,\Phi (\b r,\zeta)\,\sigma^{l\bar lZ} _\mathrm{tot}(\b r, s)=\int d^2r\, \int_0^1 d\zeta\,\Phi (\b r,\zeta)\, \int d^2B\,\im \langle i \Gamma^{l\bar lZ}(\b r, \b B, s)\rangle\,.
\eeq
This formula is the Fourier transform of the momentum space expression for the scattering amplitude.  Dipole size $\b r$ is a separation of the $l$ and $\bar l$ in the transverse configuration space  (i.e.\ in the plane perpendicular to the collision axes); this is the variable that is Fourier conjugated to the relative transverse momentum $\b k$ of the pair. We use the bold face to distinguish the transverse two-vectors.
 Impact parameter $\b B$ is a transverse vector  from the nuclear center to the   center-of-mass of the $l\bar l$ pair, see \fig{fig:nucl}. In \eq{gammap} $\Phi(\b r,\zeta)$ is the square of the photon ``wave function" averaged over the photon polarizations and summed over lepton helicities.  It  is given by \cite{dip,Nikolaev:1990ja}
\beql{phi}
\Phi(\b r, \zeta)= \frac{2\,\alpha\, m^2}{\pi}\left\{ K_1^2(rm)[\zeta^2+(1-\zeta)^2]+K_0^2(rm)\right\}\,,
\eeq
where $m$ is a lepton mass. $\zeta$ is the fraction of the photon light-cone momentum taken away by the lepton.
$\langle i\Gamma^{l\bar lZ}(\b r, \b B,s)\rangle$ is dipole--nucleus elastic scattering amplitude averaged over the proton positions (indicated by the $\langle \ldots \rangle$ symbol)
 and the dipole-nucleus cross section is, by  virtue of the optical theorem, 
\beql{llZ}
\sigma^{l\bar lZ}_\mathrm{tot}(\b r, s)=2 \int d^2B\, \im \langle i \Gamma^{l\bar lZ}(\b r, \b B , s)\rangle\,.
\eeq
Note, that in the high energy limit, the dipole cross section does not depend on $\zeta$. Integrating over $\zeta$ we get
\beql{zint}
\sigma^{\gamma Z} _\mathrm{tot}(s)=  \frac{\alpha\, m^2}{\pi}\int \frac{d^2r}{2\pi}\,\left\{ \frac{2}{3}K_1^2(rm)+K_0^2(rm)\right\}\,\sigma^{l\bar lZ}(\b r, s)\,.
\eeq

Eqs.~\eq{gammap} and \eq{phi} admit the following interpretation. At high energies, the processes of the $\gamma Z$ scattering proceeds in two separated in time stages. The first stage is fluctuation of the photon into the lepton pair $\gamma\to l+\bar l$. Let the photon, lepton and anti-lepton 4-momenta be $q=(q_+,0,0)$,  $k=(\zeta q_+, k_-,\b k)$ and $k-q= ((1-\zeta)q_+, (k-q)_-, -\b k)$, with $\zeta=k_+/q_+$. Here we use the light cone momentum notation of the 4-vector $p^\mu=(p_+,p_-,\b p)$, where  $p_{\pm}=p^0\pm p^3$. The four-scalar product is $p\cdot k= \frac{1}{2}(p_+k_-+p_-k_+)-\b p\cdot \b k$. The light-cone time span of this fluctuation is 
\beql{x-}
\Delta x_+=\frac{1}{k_-+(q-k)_--q_-}= \frac{1}{\frac{m^2+\b k^2}{\zeta q_+}+\frac{m^2+\b k^2}{(1-\zeta)q_+}}= \frac{\zeta(1-\zeta)q_+}{m^2+\b k^2}\,.
\eeq
The largest contribution to the cross section arises, as we will see later, from the longest possible $\Delta x_+$, referred to as the coherence length $l_c$ \cite{Berestetsky:1982aq}. This corresponds to $\zeta=1/2$ and $\b k=0$, so that $l_c = \frac{q_+}{4m^2}$. Obviously, $1/l_c= 4m^2/q_+$ is the minimum light-cone energy $l_-$ transfer.  

The second stage of the $\gamma Z$ process is interaction of the lepton pair  with the nucleus.  In the nucleus rest frame its time span is of the order of the target size $2R$, where $R$ is the nucleus radius. We observe that $l_c\gg 2 R$ for sufficiently high $q_+$ for a lepton of any mass $m$ supporting the physical interpretation of the two stage process. It  now becomes  clear why it is convenient to write the cross section for the process \fig{fig:gamma} in a way in which the photon ``wave function" describing the splitting process $\gamma\to l+\bar l$ is separated from the amplitude $\Gamma^{l\bar lZ}$ describing interaction of the $l\bar l$ electric dipole with the nucleus.  Indeed such an approach proved very useful for the study of the strong interactions. 

%%%%%%%%%%%%%%%%%%%%%%%%%%%%%%%%%%
\section{Glauber model for multiple scattering}\label{glauber}

%%%%%
%%%%%

 Let the nucleus quantum state be described by the  wave function $\psi_A$ that  depends on positions $\{\b b_a, z_a\}_{a=1}^Z$  of all $Z$ protons, where $\b b_a$ and $z_a$ are the transverse and the longitudinal positions  of proton $a$  correspondingly.   If the dipole--nucleus scattering amplitude 
$i\Gamma^{l\bar lZ}$ is known, then averaging over the proton positions is performed as 
\beql{av.nuc}
\langle \Gamma^{l\bar lZ}(\b r, \b B , s)\rangle\ = \int  \prod_{a=1}^Z d^2\b b_a\, dz_a\, |\psi_A(\b b_1,z_1,\b b_2,z_2,\ldots)|^2 \,\Gamma^{l\bar lZ}(\b r, \b B-\b b_1,z_1,\b B-\b b_2,z_2,\ldots, s)\,.
 \eeq
The scattering amplitude is simply related to the scattering matrix element $S$ as $\Gamma(s,\b B)= 1-S(s,\b B)$. The  later can in turn be represented in terms of the phase shift $\chi$ so that in our case 
\beql{ps}
\Gamma^{l\bar lZ}(\b r, \b B-\b b_1,z_1,\b B-\b b_2,z_2,\ldots, s)=1- \exp\{-i\chi ^{l\bar lZ}(\b r, \b B-\b b_1,z_1,\b B-\b b_2,z_2,\ldots, s)\}\,.
\eeq

At high energies, interaction of the dipole with different protons in the nucleus is independent inasmuch as the protons do not overlap in the longitudinal direction. This assumption is tantamount to taking into account only two-body interactions, while neglecting the many-body ones \cite{Glauber:1987bb}. In this approximation the phase shift $\chi^{l\bar lZ}$ in the dipole--nucleus interaction  is just a sum of the phase shifts $\chi^{l\bar lp}$ in the dipole--proton interactions i.e.\
\beq\label{phase}
\Gamma^{l\bar lZ}(\b r, \b B-\b b_1,z_1,\b B-\b b_2,z_2,\ldots, s)=1- \exp\{-i\sum_a \chi^{l\bar lp}(\b r,s,\b B-\b b_a)\}\,.
\eeq
The problem of calculating the scattering amplitude of the dipole on a system of $Z$ protons thus reduces to the problem of calculating of the scattering amplitude of the dipole on a single proton $\Gamma^{l\bar l p}$. In this approximation, $\langle e^{-i\chi} \rangle = e^{-i\langle \chi \rangle}$, i.e.\ correlations between nucleons in the impact parameter space are neglected. 

We can re-write \eq{av.nuc} and \eq{ps} as 
\beql{ps3}
\langle \Gamma^{l\bar lZ}(\b r, \b B , s)\rangle =1-\langle \prod_a[1-\Gamma^{l\bar lp}(\b r,s,\b B-\b b_a)]\rangle\,.
\eeq
For a heavy nucleus with large $Z$ we obtain
\beql{aZ}
\langle \Gamma^{l\bar lZ}(\b r, \b B , s)\rangle \approx1- \exp\{-Z\langle \Gamma^{l\bar lp}(\b r,s,\b B-\b b_a)\rangle\}
\,.
\eeq
This approximation is justified as long as $\alpha\ll 1$ and $\alpha Z\sim 1$. 
Indeed, expression in the exponent of \eq{aZ} can also be obtained by expanding 
$$
-i\chi^{l\bar lZ}=Z \ln (1-\Gamma^{l\bar l p})\approx - Z[\Gamma^{l\bar l p}-\frac{1}{2}(\Gamma^{l\bar l p})^2+\ldots ] \sim \alpha Z+\mathcal{O}(\alpha^2 Z)\,.
$$

%%%%%%%%%%%%%%%%%%%%%%%%%%%%%%%%%%
\section{Dipole--nucleus scattering amplitude}\label{scat.ampl.}

Now we turn to the calculation of the dipole--proton amplitude, see \fig{fig:ampl}. 
At high energies the $t$-channel Coulomb photons are almost real:   $l^2\approx -\b l^2$. This is the well-known Weisz\"acker-Williams approximation that we use throughout the paper.  It allows us to determine the leading logarithmic contribution to the scattering amplitude. In this approximation, the real part of the elastic dipole--proton scattering amplitude reads:
\begin{eqnarray}\label{imgamma}
\real [i\Gamma^{l\bar l p}(\b r, \b b', s)]&=&\frac{\alpha}{\pi}\int \frac{d^2l}{l^2}\left[ 
e^{i(\b b'+\frac{1}{2}\b r)\cdot \b l}- e^{i(\b b'-\frac{1}{2}\b r)\cdot \b l}\right]\nonumber\\
&=&2\alpha\int_0^\infty \frac{dl}{l} [ J_0(|\b b'+\frac{1}{2}\b r|l)-J_0(|\b b'-\frac{1}{2}\b r|l)]
\,.
\end{eqnarray} 
Taking the integral we arrive at 
\beql{llog}
\real [i\Gamma^{l\bar l p}(\b r, \b b', s)]=2\alpha \ln \frac{|\b b'-\frac{1}{2}\b r|}{|\b b'+\frac{1}{2}\b r|}\,.
\eeq
The nuclear density $\rho(\b b,z)$ is normalized such that 
\beq\label{dens.norm}
\int d^2b\, dz\,\rho(\b b,z)= Z\,.
\eeq
Therefore, the average of the real part of the amplitude over the proton position reads
\beql{a1}
\langle\real[i \Gamma^{l\bar l p}(\b r, \b B, s)]\rangle= \frac{1}{Z}\int d^2b\int_{-z_0}^{z_0} dz\, \rho\,
2\,\alpha \ln \frac{|\b b'-\frac{1}{2}\b r|}{| \b b'+\frac{1}{2}\b r|}\,,
\eeq
where $z_0=2\sqrt{R^2- b^2}$. Since $\rho\approx \mathrm{const.}$ we get
\beql{a2}
\langle \real[i \Gamma^{l\bar l p}(\b r, \b B, s)]\rangle= \frac{2\,\alpha}{Z}\int d^2b\, \rho\,2\sqrt{R^2- b^2}\,
 \ln \frac{|\b b'-\frac{1}{2}\b r|}{| \b b'+\frac{1}{2}\b r|}\,,
\eeq 
with $\b b'=\b B-\b b$, see \fig{fig:nucl}.

The imaginary part of the dipole--proton elastic scattering amplitude is depicted in \fig{fig:ampl}(b). The corresponding analytical expression reads:
\beql{re1}
\im[i\Gamma^{l\bar l p}(\b r, \b b', s)]= \frac{\alpha^2}{2\pi^2}
\int 
\frac{d^2l}{\b l^2}\int \frac{d^2l'}{\b l'^2}\left( e^{i(\b b'+\frac{1}{2}\b r)\cdot \b l}-e^{i(\b b'-\frac{1}{2}\b r)\cdot \b l}\right)
\left( e^{-i(\b b'+\frac{1}{2}\b r)\cdot \b l'}-e^{-i(\b b'-\frac{1}{2}\b r)\cdot \b l'}\right)
\,.
 \eeq
Averaging over the proton positions we derive
\begin{eqnarray}\label{re2}
\langle \im[i \Gamma^{l\bar l p}(\b r, \b B, s)]\rangle&=& \frac{1}{Z}\int d^2b\, \rho\,2\sqrt{R^2- b^2}\,\frac{\alpha^2}{2\pi^2}\nonumber\\
&&\times\int 
\frac{d^2l}{\b l^2}\int \frac{d^2l'}{\b l'^2}\left( e^{i(\b b'+\frac{1}{2}\b r)\cdot \b l}-e^{i(\b b'-\frac{1}{2}\b r)\cdot \b l}\right)
\left( e^{-i(\b b'+\frac{1}{2}\b r)\cdot \b l'}-e^{-i(\b b'-\frac{1}{2}\b r)\cdot \b l'}\right)
\,.
\end{eqnarray}

The Weisz\"acker-Williams approximation breaks down for very small momenta of the $t$-channel photons.  To determine the cutoff transverse momentum  $l_\mathrm{min}$ 
at which this approximation still holds, we  use the Ward identity applied to the lower vertex of \fig{fig:ampl}(a). It gives $p\cdot l=p_+l_-+p_-l_+=0$, where $p=(M^2/p_-,p_-,0)$ with $M$ being the proton mass. It follows that $l_+= -p_+l_-/p_-$. Thus, $-l^2= -l_+l_-+\b l^2\approx \b l^2$ if $\b l^2\gg (p_+/p_-) l_-^2$. Since the minimum value of $l_-$ is $4m^2/q_+$ (see \eq{x-} and the following text) we find
\beql{lmin}
l_\mathrm{min}=  \sqrt{\frac{p_+}{p_-}\,\frac{(2m)^4}{q_+^2}}=\sqrt{\frac{M^2(2m)^4}{p_-^2q_+^2}}= \frac{M(2m)^2}{s}\,.
\eeq
where $s=(p+q)^2\approx p_-q_+$. $l_\mathrm{min}$ corresponds to the maximum possible impact parameter $b'_\mathrm{max}\sim 1/l_\mathrm{min}$. As we will see in the next section, the energy dependence of the dipole--nucleus amplitude arises from the logarithmic dependence on these cutoff scales (this is why a precise value of the proportionality coefficient between $b'_\mathrm{max}$ and $1/l_\mathrm{min}$ is not very important).

%%%%%%%%%
\section{Bethe-Heitler limit}\label{sec:Born}

Before we consider a general case it is instructive to discuss the scattering process in the Born approximation. We would like to (i) confirm that we reproduce the result of  Bethe and Heitler \cite{Bethe:1934za} and (ii) determine the range of impact parameters $B$  that gives the largest contribution to the cross section.

The Born approximation corresponds to taking into account only the leading terms in the coupling $\alpha$. Expanding the exponent in \eq{aZ} we have
\beql{born}
\langle \Gamma^{l\bar lZ}(\b r, \b B, s)\rangle= Z\langle  \Gamma^{l\bar lp}(\b r, \b B, s)\rangle-\frac{1}{2}Z^2 \langle  \Gamma^{l\bar lp}(\b r, \b B, s)\rangle^2+\ldots \,.
\eeq
%%%%
\begin{figure}[ht]
      \includegraphics{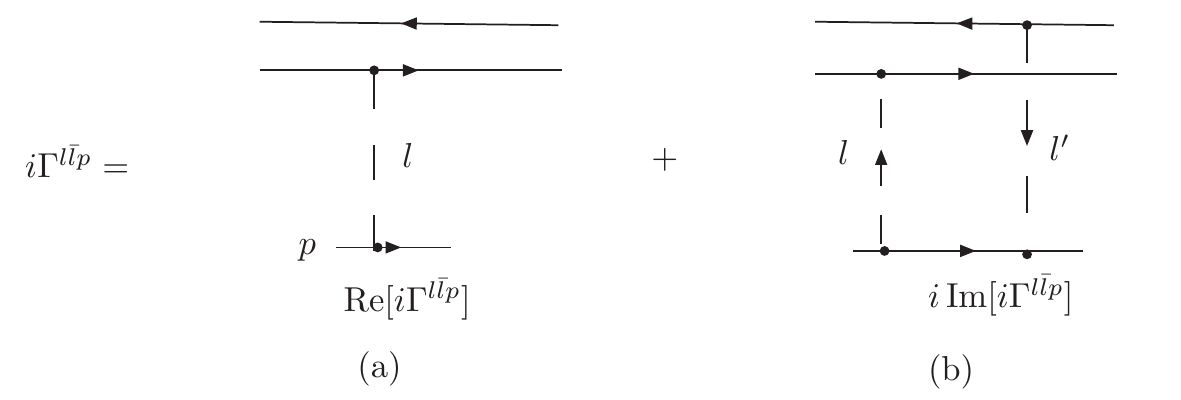} 
  \caption{The first two terms in the expansion of $\Gamma^{l\bar l p}$ in $\alpha$. }
\label{fig:ampl}
\end{figure}
%%%%%
The real part of $i\Gamma^{l\bar lp}$ stems from a diagram with a single photon exchange, while its imaginary part arises from the double photon exchange (in the high energy limit these diagrams are purely real and imaginary  correspondingly), see \fig{fig:ampl}. Therefore, at the lowest order in $\alpha$, the imaginary part of $i\Gamma^{l\bar lp}$ corresponds to the two photon exchange \fig{fig:ampl}(b) that gives -- according to \eq{born} -- contribution of order $\alpha^2 Z\sim \alpha$ to the cross section \eq{llZ}. On the other hand, the real part of $(i\Gamma^{l\bar lp})^2$ corresponds (at the leading order in $\alpha$) to the product of the two diagrams \fig{fig:ampl}(a) with single photon exchanges; it is of the order $\alpha^2 Z^2\sim 1$ \footnote{This is in sharp contrast with QCD where the single gluon exchange amplitude is identically zero because the  non-Abelian symmetry generators are traceless. Thus, there is no elastic scattering at this order. As a consequence the leading contribution in QCD is of the order of atomic weight  $A$.}. Thus, in the Born approximation
\beql{xsectB}
\sigma^{l\bar lZ}_ {\mathrm{tot},B}(\b r, s)= 2\int d^2B\, \left\{\frac{1}{2}Z^2\langle  \real[i\Gamma^{l\bar lp}(\b r, \b B, s)]\rangle^2+ Z\, \langle  \im[i\Gamma^{l\bar lp}(\b r, \b B, s)]\rangle\right\}\,.
\eeq
The reason for keeping a sub-leading term in \eq{xsectB} (the term in the integrand proportional to $Z$) is that in the nuclear medium the real part of the dipole--proton amplitude vanishes, see Eq.~\eq{z2} and then the main contribution stems from the imaginary part of that amplitude. 

%%%
\subsection{Single nucleus}\label{realistic}

Values of the  dipole--proton impact parameter $b'$ can range up to its maximal value $b'_\mathrm{max}$, which is determined as the inverse minimal momentum transfer, i.e.\ $b'_\mathrm{max}\sim 1/l_\mathrm{min}$, where $l_\mathrm{min}$ is given by \eq{lmin}. For electron production at the collision energy of $\sqrt{s}= 10$~GeV we estimate $b'_\mathrm{max}\sim  20$~nm whereas  the radius of the Gold nucleus is 7~fm. Thus for realistic nuclei  $b'_\mathrm{max}\gg R$. (In the case of $\tau$ photo-production this approximation holds for energies larger than about $\sqrt{s}= 30$~GeV). 
  
 We will demonstrate shortly, that the large logarithmic contribution to the dipole cross section \eq{xsectB} comes from the region $b'\gg b$ and $b'\gg r/2$. Because by definition  $b\le R$, 
this is equivalent to $B\approx b'\gg\max\{R,\lambda/2\}$, where $\lambda=1/m$ is the  Compton wavelength of a lepton. Here we used the fact  
 that the characteristic value of the dipole size is $r\sim 1/m$ since it corresponds to a sharp maximum of the integrand of \eq{zint}. Note, that for $e^-e^+$ dipole $R\ll \lambda\ll b'_\mathrm{max}$, while for $\mu$ and $\tau$ dipoles $\lambda\ll R\ll b'_\mathrm{max}$. 
  
Using  $b'\approx  B\gg b$, see \fig{fig:nucl}, we can write \eq{a2} as
\beql{z1}
\langle\real[i \Gamma^{e^-e^+ p}(\b r, \b B, s)]\rangle\approx  2\,\alpha\,
 \ln \frac{| \b B-\frac{1}{2}\b r|}{| \b B+\frac{1}{2}\b r|}\,
\frac{1}{Z} \int d^2b\, \rho\,2\sqrt{R^2- b^2}=
2\,\alpha\,
 \ln \frac{| \b B-\frac{1}{2}\b r|}{| \b B+\frac{1}{2}\b r|}
\,.
\eeq 
To obtain the cross section we need to integrate over the impact parameter $\b B$. Expand the logarithm in \eq{z1} at $B\gg r/2$ as
\beql{exp}
\ln \frac{| \b B-\frac{1}{2}\b r|}{| \b B+\frac{1}{2}\b r|} \approx 
 \frac{\b B\cdot \b r}{\b B^2}+\mathcal{O}(r^3/B^3)\,,
\eeq
we derive
\beql{z1a}
\langle\real[i\Gamma^{e^-e^+ p}(\b r, \b B, s)]\rangle\approx 2\,\alpha\, \frac{\b B\cdot \b r}{\b B^2}\,.
\eeq 

Turning to the imaginary part we can write  Eq.~\eq{re2} as follows
\begin{subequations}
\begin{eqnarray}
\langle \im[i \Gamma^{e^+e^- p}(\b r, \b B, s)]\rangle&=&\frac{\alpha^2}{2\pi^2}\,
\int 
\frac{d^2l}{\b l^2}\int \frac{d^2l'}{\b l'^2}\nonumber\\
&&
\times\left( e^{i(\b B+\frac{1}{2}\b r)\cdot \b l}-e^{i(\b B-\frac{1}{2}\b r)\cdot \b l}\right)
\left( e^{-i(\b B+\frac{1}{2}\b r)\cdot \b l'}-e^{-i(\b B-\frac{1}{2}\b r)\cdot \b l'}\right)
\label{relarge}\\
&=& \frac{1}{2}\,4\,\alpha^2 \ln^2 \frac{| \b B-\frac{1}{2}\b r|}{| \b B+\frac{1}{2}\b r|}
\approx 2\,\alpha^2\,\frac{(\b B\cdot \b r)^2}{\b B^4}
\,.
\label{relarge1}
\end{eqnarray}
\end{subequations}
To the leading order in $\alpha$ we can neglect  the small absorption effect given by \eq{relarge1} and write for the scattering amplitude
\beql{sa-electron}
\langle \Gamma^{l\bar l p}(\b r, \b B, s)\rangle\approx i\,2\,\alpha\, \frac{\b B\cdot \b r}{\b B^2}\,.
\eeq
Consequently, in the high energy and Born approximations the dipole--nucleus cross section becomes
\beql{xsect-light}
\sigma^{l\bar lZ}_{\mathrm{tot},B}(\b r, s)= Z^2\int d^2B\, 4\,\alpha^2\, \frac{(\b B\cdot \b r)^2}{\b B^4}
\eeq
Denote $\mathcal{R}= \max\{R,\lambda/2\}$. Integration over $B$ in the interval 
$\max\{R,r/2\}<B<b_\mathrm{max}$ gives  
\beql{bh}
\sigma^{l\bar l Z}_{\mathrm{tot},B}(\b r, s)= 4\pi \alpha^2 r^2Z^2\ln \frac{b_\mathrm{max}}{\mathcal{R}}\,,
\eeq
where the logarithmic accuracy allowed us  replacement $r=1/m$ in the argument of the logarithm. 
The total photo-production cross section off a nucleus at the order $\alpha^3$ in the perturbation theory reads:
\begin{eqnarray}\label{ii1}
\sigma^{\gamma Z}_{\mathrm{tot},B}(s)&=&\sum_l \int \frac{d^2r}{2\pi} \int_0^1 d\zeta\, \Phi(\b r, \zeta)\frac{1}{2}\,\sigma^{l\bar lZ}_{\mathrm{tot},B}(\b r, s)\nonumber\\
&=&
\sum_l  4\alpha^3Z^2 m_l^2\int_0^\infty dr r\left[ \frac{2}{3}K_1^2(rm_l)+K_0^2(rm_l)\right]r^2\ln \frac{b_\mathrm{max}}{\mathcal{R}}\,,
\end{eqnarray}
where the sum runs over all lepton species. Taking integrals over the dipole sizes finally produces 
\beql{gZLO}
\sigma^{\gamma Z}_{\mathrm{tot},B}(s)=  \sum_l 4\alpha^3Z^2 m_l^2\left( \frac{2}{3}\frac{2}{3m_l^4}+\frac{1}{3m_l^4}\right)\ln \frac{b_\mathrm{max}}{\mathcal{R}}\,.
\eeq
In particular, for electrons
\beql{elec-fino}
\sigma^{\gamma Z\to e^+e^-X}_{\mathrm{tot},B}(s)= \frac{28}{9}\frac{\alpha^3Z^2}{m^2_e}\ln \frac{s}{2Mm_e}\,.
\eeq
In a frame where nuclear proton moves with velocity $v= (p_+-p_-)/(p_++p_-)$ the minus component of its four-momentum is $p_-= M\sqrt{(1-v)/(1+v)}$. Therefore, 
$s=2M\omega\sqrt{(1-v)/(1+v)}$,  where the incoming photon frequency is $\omega = q_+/2$. Eq.~\eq{elec-fino} now reads
\beql{gZLO-2}
\sigma^{\gamma Z\to e^+e^-X}_{\mathrm{tot},B}(s)=  \frac{28}{9}\frac{\alpha^3Z^2}{m_e^2}\ln \left(\frac{\omega}{m_e}\sqrt{\frac{1-v}{1+v}}\right)\,.
\eeq
This is the high energy limit of the formula derived by Bethe and Heitler \cite{Bethe:1934za}.  In the Appendix we discuss the energy-independent correction to this equation.

%%%%%
\subsection{Nuclear medium}\label{sec:med}

Consider a nuclear matter  having an arbitrary large number of protons $Z$. We assume  that the inter-proton distance is much smaller than the Compton wavelength of electron, i.e.\  $\rho^{-1/3}\ll \lambdabar$. This allows treating the nuclear matter as the continuous medium of  density $\rho$.
Let the medium  transverse size be $d\gg b'_\mathrm{max}$ and longitudinal size $L\ll l_c$. In the medium rest frame $l_c\simeq b'_\mathrm{max}$\footnote{The values of $b'_\mathrm{max}$ 
at  $\sqrt{s}=10$~GeV for $e$, $\mu$ and $\tau$ are  20~nm,  0.4~pm,  and  2~fm correspondingly.\label{foot}}. Therefore, the nuclear medium has a form of a film with $d\gg L$. What does the lepton photo-production look like in this case? The key observation in this case is that the dipole--proton interaction range is much smaller compared to the transverse size of the medium. 
Therefore, $b'\ll B$ and  then it follows from \eq{a2} that the real part of the elastic dipole--proton scattering amplitude  vanishes. Indeed,  changing the integration variable  $\b b\to \b b'$ in \eq{a2} and taking the limit of small $\b b'$  we obtain
\begin{eqnarray}\label{z2}
\langle \real[i\Gamma^{l\bar l p}(\b r, \b B, s)]\rangle&=&\frac{2\,\alpha}{Z}
\int d^2b' \rho(\b B-\b b')\,L\,
 \ln \frac{| \b b'-\frac{1}{2}\b r|}{| \b b'+\frac{1}{2}\b r|}\nonumber\\
 &\approx&
 \frac{2\,\alpha}{Z}
 \rho(B)\,L\,\int d^2b'\,
 \ln \frac{| \b b'-\frac{1}{2}\b r|}{| \b b'+\frac{1}{2}\b r|}=0\,.
\end{eqnarray}
 The last equation follows because the integrand is an odd function under the reflection $\b b'\to -\b b'$. The first non-vanishing contribution to the real part of the amplitude is proportional to the nuclear density gradient, which has the largest value at the  boundary. These small effects can be taken into account if the nuclear density distribution is known.  

The leading contribution arises from the imaginary part of the amplitude that reads
\begin{subequations}
\begin{eqnarray}
\langle \im[i \Gamma^{l\bar l p}(\b r, \b B, s)]\rangle&\approx&\frac{1}{Z}
\rho\,L\,\int d^2b'\, \frac{\alpha^2}{2\pi^2}\,\nonumber\\
&\times&\int 
\frac{d^2l}{\b l^2}\int \frac{d^2l'}{\b l'^2}\left( e^{i(\b b'+\frac{1}{2}\b r)\cdot \b l}-e^{i(\b b'-\frac{1}{2}\b r)\cdot \b l}\right)
\left( e^{-i(\b b'+\frac{1}{2}\b r)\cdot \b l'}-e^{-i(\b b'-\frac{1}{2}\b r)\cdot \b l'}\right)
\label{resmall}\\
&=&\frac{1}{Z} \rho\,L\, 4\,\alpha^2\int \frac{d^2l}{\b l^4}\,
\left( 1-e^{i\b r\cdot \b l}\right)
\label{resmall2}\\
&\approx& \frac{1}{Z} \rho\,L\, 4\,\alpha^2\,(2\pi)\,\frac{1}{4}\,r^2
\ln\frac{2}{rl_\mathrm{min}}\,,\label{resmall3}
\end{eqnarray}
\end{subequations}
where in the last line we kept only the logarithmically enhanced term. 
We see that in the case of the nuclear medium, the entire elastic dipole--proton amplitude is approximately imaginary and is given by Eq.~\eq{resmall3}. Moreover, unlike \eq{relarge1} it increases logarithmically with energy. 
Substituting \eq{z2} and \eq{resmall3} into \eq{xsectB} we obtain in the Born approximation
\beql{bh2}
 \sigma^{l\bar l Z}_{\mathrm{tot},B}(\b r, s)=4\pi Z\,\alpha^2\,r^2\ln\frac{2}{rl_\mathrm{min}}\,.
 \eeq
Note, that cross section \eq{bh2}  is $Z$ times smaller then the one given by \eq{bh}.

%%%%%%%%%%%%%%%%%%%%%%%%%%%%%%%%%%%%%
\section{Unitarity effects}\label{resum}

The unitarity relation applied to the elastic scattering amplitude at a given impact parameter reads (we suppress the argument $(\b r, \b b, s)$ of all functions)
\beql{unitarity}
2\,\im(i \Gamma^{l\bar lZ})= | \Gamma^{l\bar lZ}|^2+G^{l\bar lZ}\,,
\eeq
where $G^{l\bar lZ}$ is the inelastic scattering amplitude. Using \eq{aZ} we can solve \eq{unitarity} as
\beql{gin}
G^{l\bar lZ}= 1-e^{-Z \langle \im(i\Gamma^{l\bar l p})\rangle } \,.
\eeq
It follows that the total, inelastic and elastic cross sections are given by
\begin{subequations}\label{secs}
\begin{eqnarray}
\sigma_\mathrm{tot}^{l\bar l Z}&=& 2\int d^2B \left\{ 1-  \cos[ Z\langle\real(i\Gamma^{l\bar l p})\rangle]\, e^{ -Z \langle \im(i\Gamma^{l\bar l p})\rangle}\right\} \\
\sigma_\mathrm{in}^{l\bar l Z}&=& \int d^2B \left\{ 1-e^{-Z \langle \im(i\Gamma^{l\bar l p})\rangle }\right\}\\
\sigma_\mathrm{el}^{l \bar l Z}&=& \int d^2B \left\{ 1-2\cos[ Z\langle\real(i\Gamma^{l\bar l p})\rangle]\,e^{ -Z \langle \im(i\Gamma^{l\bar l p})\rangle}+e^{-Z \langle\im(i\Gamma^{l\bar l p})\rangle } 
\right\}\,.
\end{eqnarray}
\end{subequations}

When $Z \langle\Gamma^{l\bar l p}\rangle\sim 1$ deviation from the Born approximation becomes large and the unitarity effects set in.  We proceed to analyze the unitarity corrections in two extreme cases. 

%%%%
\subsection{Single nucleus}\label{xsec:rn}

In the case of photon--nucleus scattering, we derive using \eq{sa-electron} and integrating over the directions of $\b B$:
\beql{xxx1}
\sigma_\mathrm{tot}^{l\bar l Z}\approx  \sigma_\mathrm{el}^{e^- e^+ Z}= 
2\int d^2B\left( 1- \cos\{ 2\,Z\,\alpha\,\frac{\b B\cdot \b r}{B^2}\}\right)= 
4\pi \int_0^{B_\mathrm{max}}dB\, B\, [ 1- J_0(2Z\alpha r/B)]
\, .
\eeq
Introducing the dimensionless variable $x=2Z\alpha r/(2B_\mathrm{max})$ we can express this integral in terms of the generalized hypergeometric function:
\beql{final-ee}
\sigma_\mathrm{tot}^{l\bar l Z}(\b r, s)=\frac{(2Z\alpha r)^2\pi}{8}\{ x^2\, _{2}F_3[(1,1),(2,3,3),-x^2]+4(2-2\gamma-\ln x^2) \}\,.
\eeq
where $\gamma$ is Euler's constant.
The total photo-production cross section of light leptons is obtained using \eq{zint}. Actually, it is convenient to directly  plug \eq{xxx1} into \eq{zint} and first integrate over $r$  and then over $B$. The result can be expressed in terms of another dimensionless variable  $y= \alpha Z/(B_\mathrm{max}m)$ as:
\begin{eqnarray}\label{gammaZ-ee}
\sigma^{\gamma Z}_\mathrm{tot}(s)&=& \frac{\alpha^3 Z^2}{9m^2y^3}
\bigg( 2y -28 y^3\ln(2y)\nonumber\\
&&
+\{-3\sqrt{1+y^2}+24y^2\sqrt{1+y^2}+\cosh[3\ln(y+\sqrt{1+y^2})]\}\,\ln(y+\sqrt{1+y^2})\bigg)
\,.
\end{eqnarray}
To find the high energy asymptotic we expand \eq{gammaZ-ee}  at small $y$ and find 
\beql{v2}
\sigma^{\gamma Z}_\mathrm{tot}(s)= \frac{28\alpha^3 Z^2}{9m^2}\big[\ln \frac{1}{2y}+\frac{41}{42}+\frac{12}{35}y^2+\cdots\big]\,.
\eeq
In the leading logarithmic approximation that we use in this  paper, only the leading logarithm in the square brackets of \eq{v2} can be guaranteed. Indeed, it reproduces 
the Bethe-Heitler formula \eq{elec-fino}. However, Eq.~\eq{v2} does allow us to conclude that corrections to the leading logarithm must not increase with energy. 
Therefore,  in the case of lepton pair photo-production on a single nucleus the  unitarity corrections do not grow with energy at the leading order in $\alpha\ll 1$ and $\alpha Z\sim 1$. Actually, the Glauber model that we use in this paper also allows us to reproduce the subleading (i.e.\ energy-independent) corrections to Eq.~\eq{elec-fino} that were first obtained by Bethe and Maximon \cite{Bethe:1954zz}.  This result can be easily derived using the formalism developed by Ivanov and Melnikov in \cite{Ivanov:1998ka}. The derivation is outlined in the Appendix.  
In the framework of our approach it is straightforward to calculate corrections of order $\alpha^2Z$ to the Bethe-Maximon formula by taking into account the imaginary part of the amplitude \eq{relarge1}. This problem will be addressed elsewhere.

%%%
\subsection{Nuclear medium}\label{sec:med.sec}

Conclusions of the previous subsection are dramatically reversed in the case of scattering off a nuclear medium. 
For the total dipole cross section we obtain after substituting \eq{z2} and \eq{resmall3} into \eq{secs}
\beq\label{sh}
\sigma_\mathrm{tot}^{l\bar lZ}(\b r, s)= 2\pi d^2 \,(1-\exp\{-\rho\, L\, 2\pi \alpha^2 r^2\ln \frac{2}{rl_\mathrm{min}} \})\,,
\eeq
Similar results were obtained for an infinite medium in \cite{Lyuboshits:1981ky}.
Substituting into \eq{zint} we derive
\beql{gammaZ-mm}
\sigma^{\gamma Z}_\mathrm{tot}(s)= 2\alpha\, m^2\,d^2\int \frac{d^2r}{2\pi}\, \left\{ \frac{2}{3}K_1^2(rm)+K_0^2(rm)\right\}\,\left[ 1-e^{-\rho\, L\, 2\pi \alpha^2 r^2\ln \frac{2}{rl_\mathrm{min}} }\right]\,.
\eeq
At large $Z$ the multi-photon effects start to play an important role. 

\begin{figure}[ht]
      \includegraphics[width=8cm]{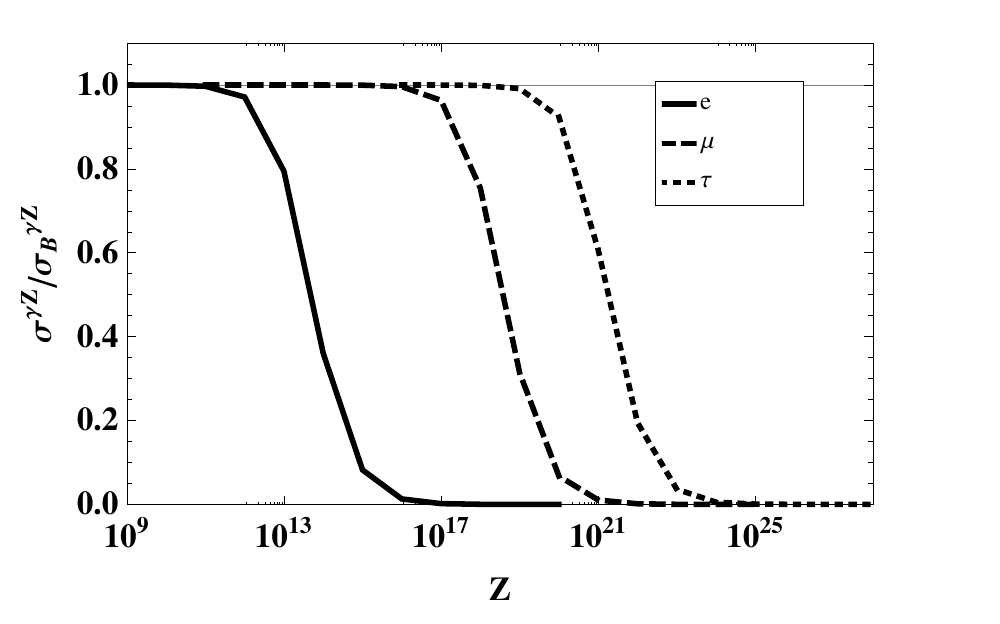} 
  \caption{Ratio of the lepton photo-production cross section and its Born approximation as a function of the ``nucleus" charge $Z$. $\sqrt{s}=10$~GeV, $L$=1nm, and $d$=100~nm.}
\label{fig:satur}
\end{figure}

It is instructive to study the unitarity corrections by  increasing the proton number $Z$ while keeping the medium size given by  $L$ and $d$ fixed. This corresponds to an increase in the proton number density. We are interested to know at what $Z$'s the multi-photon effects become observable.  In \fig{fig:satur} we show the ratio of the total photo-production cross  calculation to its Born approximation for $\sqrt{s}=10$~GeV and $L=1$~nm, $d=100$~nm. We observe that for electrons the deviation from the linear Born regime starts at about $Z\sim 10^{12}$ charges. At $Z\sim 10^{13}$ the cross section is essentially black (i.e.\ $Z$-independent).

%We are interested to know at what charge $Z$ of the ``nucleus"  would the lepton photo-production show a deviation from the Born approximation. The proton density is given by $\rho= Z/(\frac{4}{3}\pi R^3)$. The nuclear radius is $R=1.2 A^{1/3}$~fm. Requirement that $\Omega=1$ yields the minimal ``nuclear" radius at which the unitarity effects become significant
%\beql{RU}
%R_\mathrm{U}=\frac{2\, m^2 (1.2\,\mathrm{fm})^3}{3\,\alpha^2 \ln \frac{2s}{Mm}}\,,
%\eeq
 %where we assumed that $A=2Z$.  The onset of unitarity effects at  $\sqrt{s}=10$~GeV for electrons happens at the ``nuclear" radius of $R_\mathrm{U}= 8\cdot 10^{-5}$~fm.  For muons it happens  
% at $R_\mathrm{U}= 7$~fm and for $\tau$ only at $R_\mathrm{U}= 2.9\cdot 10^3$~fm. 
%Using the values for $b'_\mathrm{max}$ given in the footnote~\ref{foot} we see that once the condition $R\gg b'_\mathrm{max}$ is satisfied, the cross sections for the electron and muon photo-production are black. 

%%%
%%%%%

%%%%%%%%%%%%%%%%%%%%%%%%%%%%%%%%
\section{Discussion}

In this article we discussed an effect of  multiple scattering on the high energy lepton photo-production off a heavy nucleus and nuclear medium. The heavy nucleus serves as a source of strong Coulomb field. Resummation of multiple scattering of a lepton dipole in the nucleus amounts to resummation of diagrams with an arbitrary number of photons attached to the lepton loop. In the heavy nucleus case we selected only diagrams that are enhanced by the power of $Z$ and resummed them using the Glauber model. We found that   dominance of elastic scattering leads to energy-independent  unitarity corrections. These corrections do not saturate the geometric limit of the cross sections. 
 The unitarity corrections that grow with energy can appear as a result of multiple lepton pair production at very high energies \cite{Gribov:1970ik}. The dipole model that we used in this paper is a  convenient tool for taking into account such high energy evolution effects. We plan to address this problem elsewhere.

Lepton photo-production off a nuclear medium is different in many aspects. Here the processes are predominantly inelastic. At very large $Z$ the cross sections \eq{secs} saturate at the geometric limit due to multi-photon interactions with the medium.  An additional contribution to the saturation stems from the multiple  electron-positron pair production \cite{Gribov:1970ik,Mueller:1988ju}. This leads to  growth of  the imaginary part of $i\Gamma^{l\bar lp}$ as a power of energy  $s^\Delta$ with $\Delta\approx (11/32)\pi \alpha^2$\cite{Mueller:1988ju}. However, since $\Delta$ is very small this effect is mostly of academic interest. Parametrically, lepton photo-production  off the nuclear medium is very similar  to the hadron photo-production in strong interactions. The real part of $i\Gamma^{l\bar lp}$ vanishes identically in that case because it is proportional to the vanishing trace of the Gell-Mann matrices.  In QCD, the quark--anti-quark color dipole interacts with the heavy nucleus by means of multiple exchange of virtual gluons. Additionally, multiple gluon emission contributes to the power-law dependence of the cross section. The corresponding power  is not small leading to important observable effects. Unitarity correction in QCD grow very large at high energy leading to saturation of the scattering amplitudes \cite{Gribov:1983tu,McLerran:1993ni,Kovchegov:1996ty}. In lepton photo-production
 QCD evolution will arise at high orders of perturbation theory. 

 In the case of the nuclear medium, weakness of the EM coupling can be compensated by large number density of charges. In  a more realistic model one  should also take into account screening of the Coulomb fields by electrons leading to the emergence of the nuclear form-factor. However, our emphasis in this article is the study of electro-magnetic interactions  in a high density system. A high density of charges can be realized also  in ultra-fast pulse  lasers that emit pulses containing as many as $N\sim 10^{20} $ photons and having dimensions $L<d$. In this case the inelastic in-medium effects start to play a significant role in large range of collision energies.   
Although fluctuation of  laser photons into electric dipoles is suppressed by the power of $\alpha$, the created Coulomb field is enhanced by  large effective charge $Z$. We are going to address this intriguing problem in a forthcoming publication.

%%%%%%%%%%%%%%%%%%%%%%%%%%%%%%%%
\acknowledgments
I  am grateful to Genya Levin and Yura Kovchegov for reading a draft version of this article and making many useful comments. I would like to thank Dmitry Ivanov for important comments and useful references. 
This work  was supported in part by the U.S. Department of Energy under Grant No.\ DE-FG02-87ER40371. I 
thank RIKEN, BNL, and the U.S. Department of Energy (Contract No.\ DE-AC02-98CH10886) for providing facilities essential
for the completion of this work.

%%%%%%%%%%%%%%%%%%%%
\appendix
\section{Derivation of the Bethe-Maximon formula}

Eq.~\eq{xxx1} is written in the leading logarithmic approximation that does not allow one to reproduce the subleading energy-independent terms in the total cross section.  However, the Glauber model allows one to calculate the total cross section with better accuracy. To this end, one has to abandon the approximation of \eq{exp} and use formula \eq{z1} for the real part of the amplitude  in  \eq{secs}. We have
\beql{app1}
\sigma_\mathrm{tot}^{\gamma Z}(s)= \frac{\alpha m^2}{\pi}
\int \frac{d^2r }{2\pi}\big[ \frac{2}{3}K_1^2(mr)+K_0^2(mr)\big]\,2\int d^2B\bigg\{ 
1-\cos\big[2\alpha Z\ln \frac{|\b B-\b r/2|}{|\b B+\b r/2|}\big]\bigg\}\,.
\eeq
Introducing a new integration variable $\b R= \b B+\b r/2$ we re-write \eq{app1} as
\beql{app3}
\sigma_\mathrm{tot}^{\gamma Z}(s)= \frac{\alpha m^2}{\pi}
\int \frac{d^2r }{2\pi}\big[ \frac{2}{3}K_1^2(mr)+K_0^2(mr)\big]\,2\int d^2 R
\bigg\{ 1-\frac{1}{2}\bigg( \frac{|\b R-\b r|}{R}  \bigg)^{2i\alpha Z} - 
\frac{1}{2}\bigg( \frac{|\b R-\b r|}{R}  \bigg)^{-2i\alpha Z}
\bigg\}
\eeq
Consider 
\beql{app5}
\Delta \sigma_\mathrm{tot}^{\gamma Z}= \sigma_{\mathrm{tot}}^{\gamma Z}-\sigma_{\mathrm{tot},B}^{\gamma Z}\,,
\eeq
i.e.\ the contribution due to multiple rescatterings. The term corresponding to the Born approximation is obtained by expanding the expression in the curly brackets in \eq{app3} at small $\alpha Z$. One then arrives at the formula derived by Ivanov and Melnikov \cite{Ivanov:1998ka}. It reads:
\begin{eqnarray}\label{app7}
\Delta \sigma_\mathrm{tot}^{\gamma Z}&=&2 \frac{\alpha m^2}{\pi}
\int \frac{d^2r }{2\pi}\big[ \frac{2}{3}K_1^2(mr)+K_0^2(mr)\big]\nonumber \\
&&
\times \int d^2 R
\bigg\{ 1-\frac{1}{2}\bigg( \frac{|\b R-\b r|}{R}  \bigg)^{2i\alpha Z} - 
\frac{1}{2}\bigg( \frac{|\b R-\b r|}{R}  \bigg)^{-2i\alpha Z}
-4 (\alpha Z)^2 \ln^2\frac{|\b R-\b r|}{R}  \bigg\}\,.
\end{eqnarray}
Taking integrals  over $\b R$ and $\b r$ yields:
\beql{app9}
\Delta \sigma_\mathrm{tot}^{\gamma Z}=-\frac{28}{9m^2}\alpha^3Z^2 
\frac{1}{2}[\psi(1-i\alpha Z)+\psi(1+i \alpha Z)-2\psi(1)]\,,
\eeq
where $\psi(z)$ is the logarithmic derivative of the Gamma function. As expected, the unitarity corrections are independent of energy in agreement with the results of  \sec{xsec:rn}. 

To calculate the energy-independent term of the Born approximation (the last term in the curly brackets of \eq{app7}) one needs to take into account the fact that the smallest longitudinal momentum transfer depends on $\zeta$ and $\b k$, see \eq{x-}. In the logarithmic approximation, we maximized \eq{x-} with respect to both $\zeta$ and $\b k$. Now it must be retained which means that {\tt (i)} the regularization of the logarithmically divergent 
integrals must be done in momentum space and {\tt (ii)} the $\zeta$ integral (see \eq{gammap})  must be done after the regularization since the cutoff depends on it. We refer the interested reader to the Ref.~\cite{Ivanov:1998ka} for details. The result is  
\beql{app11}
\sigma_{\mathrm{tot},B}^{\gamma Z}(s)= \frac{28}{9}\frac{\alpha^3Z^2}{m^2_e}\bigg(\ln \frac{s}{Mm_e}  -\frac{109}{42}\bigg) \,.
\eeq
The second term in the brackets is the subleading correction we were looking for. The final result is the sum of \eq{app9} and \eq{app11} in agreement with \cite{Bethe:1954zz}.

%%%%%%%%%%%%%%%%%%%%%%%%%%%%%%%%%%%%%

\end{document}